\documentclass[prb,aps,twocolumn,a4]{revtex4}
\usepackage{rotate,bm,epsfig}
\newcommand{\be}{\begin{equation}}
\newcommand{\ee}{\end{equation}}
\newcommand{\ber}{\begin{eqnarray}}
\newcommand{\eer}{\end{eqnarray}}

\newcommand{\dmdt}{\frac{{\rm d}\bm{m}}{{\rm d}t}}
\newcommand{\ud}{\left(\bm{u}\bm{\nabla}\right)}
\newcommand{\heff}{\bm{H}_{\rm eff}}
\begin{document}
\title{Current-induced magnetic vortex core switching in a Permalloy nanodisk} 
\author{Y. Liu}
\affiliation{Institut f\"ur Festk\"orperforschung IFF-9 "Elektronische
  Eigenschaften", Forschungszentrum J\"ulich GmbH, D-52425 J\"ulich,
  Germany}
\affiliation{Department of Physics, Tongji University, 200092 Shanghai, P.R. China}
\author{S. Gliga}
\affiliation{Institut f\"ur Festk\"orperforschung IFF-9 "Elektronische
  Eigenschaften", Forschungszentrum J\"ulich GmbH, D-52425 J\"ulich,
  Germany}
\author{R. Hertel}
\affiliation{Institut f\"ur Festk\"orperforschung IFF-9 "Elektronische
  Eigenschaften", Forschungszentrum J\"ulich GmbH, D-52425 J\"ulich,
  Germany}
\thanks{Author to whom correspondence should be addressed; r.hertel@fz-juelich.de}
\author{C.M.~Schneider}
\affiliation{Institut f\"ur Festk\"orperforschung IFF-9 "Elektronische
  Eigenschaften", Forschungszentrum J\"ulich GmbH, D-52425 J\"ulich,
  Germany}
\date{\today}
\begin{abstract}
We report on the switching of a magnetic vortex core in a sub-micron
 Permalloy disk, induced by a short current pulse applied in the film
 plane. Micromagnetic simulations including the adiabatic and
 non-adiabatic spin-torque terms are used to investigate the
 current-driven  magnetization dynamics. We predict that a core
 reversal can be triggered by current bursts a tenth of a nanosecond
 long. The vortex core reversal process is found to be the same as
 when an external field pulse is applied. The control of a vortex
 core's orientation using current pulses introduces the
 technologically relevant possibility to address individual
 nanomagnets within dense arrays.      
\end{abstract}
\maketitle

Patterned soft-magnetic thin film elements with lateral extension
above a critical size naturally form flux-closure
patterns \cite{Cowburn99}, which contain magnetic vortices, {\em
  i.e.}, small regions where the  magnetization direction circulates
in the film plane around a nanometer-sized core \cite{Miltat02b}. At
the center of this core, the magnetization points perpendicular to the
plane \cite{Shinjo00}. In the search for smaller and faster devices,
such magnetic patterns  have mostly been avoided favoring samples with 
uniform magnetization \cite{Gerrits02}. The study of the static and
dynamic properties of vortices and vortex cores in particular has,
however, recently emerged as a dynamic field of investigation
\cite{Choe04,Park03b,Wachowiak02,Stoll_n06}. Magnetic vortex cores
have indeed recently been considered as possible candidates for
magnetic data storage \cite{Hollinger03} due to several attractive
properties:  Their size is of only a few tens of nanometers
\cite{Wachowiak02}, they exhibit perfect bi-stability (pointing either 
upwards or downwards, defining the vortex polarization), they form
naturally and ultimately have a high thermal stability. All these
features represent important prerequisites for possible applications
to data storage. In order to store information using a vortex core,
mechanisms for a controlled switching of its orientation are required.        
The direct reversal of vortex cores by means of an applied field 
oriented perpendicular to the plane requires large field values in the
order of 500 mT \cite{Okuno02}, which gives an idea of the high
stability of these structures. Recently, it has been shown that the
vortex core could be switched by means of low fields applied in the
plane of the sample \cite{Stoll_n06}. Experimentally, it has been
shown that a sinusoidal in-plane field pulse as low as 2\,mT could be
used to reverse the core of a vortex in a sample excited at the
gyrotropic frequency, where the vortex is brought into a stationary
orbit \cite{Stoll_n06}. However, the reversal process takes a few
nanoseconds, due to the fact that the steady gyrotropic orbit motion 
is in the order of 100\,MHz \cite{Argyle84s}. Although the gyrotropic
frequency depends on the sample size \cite{Choe04}, this dependence is
weak and this frequency is practically always in the sub-GHz range, so
that switching speeds well below 1~ns are not possible using this
resonant scheme. Moreover, the core switching only occurs if the
frequency of the sinusoidal field is within a narrow range close to
the gyrotropic frequency \cite{Faehnle07}. Based on micromagnetic
simulations, an ultra-fast  
core reversal mechanism has been proposed recently \cite{Hertel07cm},
which is initiated by applying a suitably shaped unipolar in-plane
magnetic field pulse only a few picoseconds long. The required
amplitude of the field pulse is larger (ca.~70\,~mT) than in the case
of resonant switching, but this switching scheme is much faster. 
In spite of their respective advantages (low fields, high speed) both
the resonant and the non-resonant switching modes exhibit a common
problem in terms of applicability: The lack of selectivity of the
individual elements. Reliably addressing a single nanodisk inside a
dense array is very difficult using external fields.  
In this letter, we present a fast and simple method to switch magnetic
vortex cores by applying short {\em electric} current pulses, only one
hundred picoseconds long. The electric current pulse is applied in
the plane of the element. We thus show that a fast toggle core
switching mechanism can be triggered in a relatively  simple way which
is compatible with integrated circuits, thereby solving the issue of
selectivity.  

\begin{figure*}[ht]
\centerline{\epsfig{file=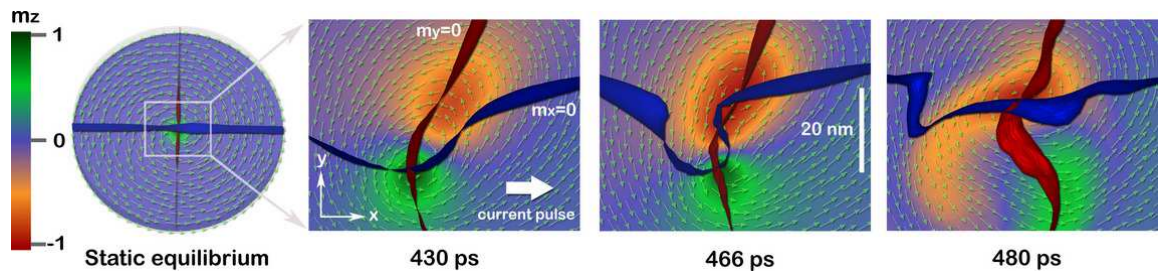,width=.85\linewidth}}
\caption{\label{scheme}
Current-induced vortex core reversal in a Py
  nanodisk of 200 nm in diameter and 20 nm thickness. A Gaussian
  current pulse is applied in the sample plane with a strength of
  5.2$\cdot10^{12}$ A/m$^2$ and a width $\sigma=100$ ps. 
  On the left, the magnetic structure of the whole sample is shown 
  at equilibrium. In the three frames on the right, the evolution of
  the magnetization is shown for a small region around the vortex
  core, where a new vortex-antivortex pair nucleates. As the vortex
  is shifted in the direction of the electron flow prior to the pair
  creation, these frames are not taken in the same areas of the
  sample. The green arrows represent the {\it x} and {\it y}
  components of the magnetization, while the cross-sections of the
  bottom surface of the sample are shown for the {\it z} component of
  the magnetization. The green area represents $m_z=+1$, whereas
  $m_z=-1$ in the orange area. The blue and red ribbons represent the
  $m_x=0$ and $m_y=0$ isosurfaces, respectively.
}  
\end{figure*}

The results were obtained using micromagnetic finite-element simulations
based on the Landau-Lifshitz-Gilbert equation. Our micromagnetic
code used in previous simulations \cite{Hertel04} has been extended in
order to consider the spin torque exerted by an electric current flowing
through the sample. This effect of current-driven magnetization
dynamics is modelled by including the adiabatic and the non-adiabatic
spin torque terms \cite{Zhang04,Thiaville05} into the Gilbert
equation:   
\be\label{orig}
\dmdt=-\gamma\bm{m}\times\heff+\alpha\bm{m}\times\dmdt-\ud\bm{m}+\beta\bm{m}\times
\left[\ud\bm{m}\right],
\ee
where $\bm{m}=\bm{M}/M_{\rm s}$ is the normalized local magnetization
($M_{\rm s}$: saturation magnetization), $\bm{H}_{\rm eff}$ is the
effective field containing the exchange and the dipolar field, $\beta$
is a dimensionless parameter describing the strength of the 
non-adiabatic term, and $\alpha$ is the Gilbert damping constant. The
vector $\bm{u}$ is pointing parallel to the electron flow direction
and has an amplitude of $u=jPg\mu_{\rm B}/(2eM_{\rm s})$
\cite{Thiaville05}, where $j$ is the current density, $P$ is the
degree of electron polarization, $g$ is the Land\'e splitting factor,
$\mu_{\rm B}$ is the Bohr magneton, and $e$ is the electron charge. 
As a model system, we consider a disk-shaped Permalloy (Py) sample of
$d=200$\,nm diameter and $t=20$\,nm thickness, as used in Ref.~\cite{Hertel07cm}.
The parameters used in the simulation are $A=13$~pJ/m (exchange
constant), $\mu_0M_{\rm s}=1.0$\,T, $\alpha=0.03$, $\beta=0.02$, $K$=0
(anisotropy constant). The sample is discretized into ca.~\,216,000 irregular 
tetrahedral elements, corresponding to a cell size of about 3\,nm. A
homogeneous current density distribution is assumed, as was done
recently in a similar study by Kasai {\em et al.}~\cite{Kasai06}, where
the resonant, current-pulse induced gyrotropic vortex core motion was
investigated.  

The current-induced vortex core reversal is studied for short
Gaussian-shaped current pulses ($\sigma=100$\,ps) of varying
strengths. To obtain a core reversal, we found that for the considered
sample and for 100 ps pulse duration, the pulse amplitude must exceed a
minimum value of $j$=4.7$\cdot 10^{12}$A$/$m$^2$. Although such a high
current density might endanger the structural stability of the sample
if it was applied continuously, the damaging effects of the current
should be small in the present case where only ultrashort pulses are
used.

A typical example of the simulated vortex core reversal process is
shown in Fig.~1, starting with a vortex whose core is
pointing in the positive $z$-direction. 
The micromagnetic processes leading to the vortex core reversal are
identical to the ones described in Ref.~\cite{Hertel07cm}, where field
pulses were used to trigger a vortex core switching. The isosurface
representation introduced in Ref.~\cite{Hertel06} has been 
used to precisely locate the vortex core. It can be seen that the
in-plane magnetization is first heavily distorted as a result of the
applied current, which is running through the sample. After
ca.~\,460\,ps the distortion eventually leads to the creation of a 
vortex-antivortex pair, which can unambiguously be recognized by the
two additional crossings of the $m_x=0$ and $m_y=0$ isosurfaces. Both
cores of the new pair are pointing in the opposite $z$-direction of
the initial core \cite{Hertel07cm}. The newly formed antivortex and
the oppositely polarized initial vortex subsequently annihilate as
described in Ref.~\cite{Hertel06}.  The latter subprocess unfolds over
approximately 10 ps and leaves a single vortex core, which is
oppositely polarized with respect to the initial one. The formation of
vortex-antivortex pairs after application of short current pulses is
consistent with recent experimental observations by Kl\"aui {\it et
  al.} \cite{Klaui06} on the domain wall mobility in thin magnetic
strips.    
\begin{figure}[h]
\centerline{\epsfig{file=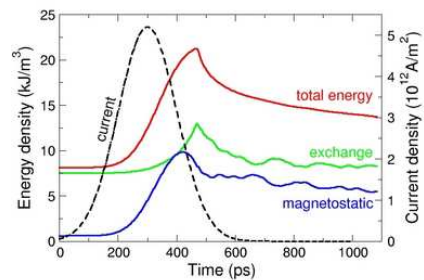,width=.7\linewidth}}
\caption{\label{plots1}
Evolution as a function of time of the total, 
magnetostatic and exchange energy densities in the Py disk following a
current pulse of a duration of  
$\sigma=100$\,ps and a peak value of 5.2$\cdot10^{12}$ A/m$^2$.
}  
\end{figure}

Fig.~2 shows the evolution in time of the sample's spatially averaged
energy density in the case of an applied current pulse of 
5.2$\cdot10^{12}$ A/m$^2$ and $\sigma=100$\,ps. After the core
switches, the total energy rapidly decreases as a consequence of the 
annihilation of the vortex-antivortex pair. It can clearly be seen
that the exchange energy decreases after the core switching event,  
in agreement with the interpretation given in Ref.~\cite{Hertel07cm}
that it is the exchange field which is ultimately driving the core reversal.   

\begin{figure}[h]
\centerline{\epsfig{file=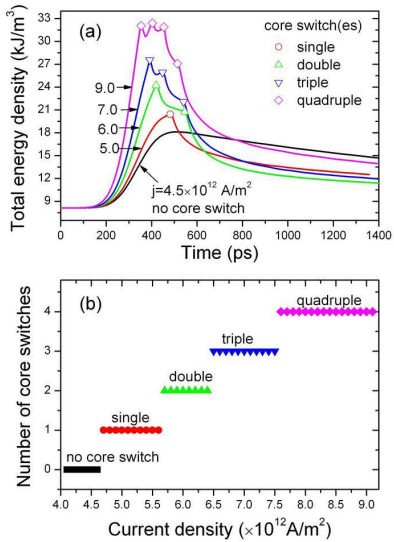,width=.65\linewidth}}
\caption{\label{plots2}
(a) Total energy for multiple core switches.
(b) Number of times the vortex core switches as a function of the 
applied current.
}
\end{figure}

By increasing the pulse strength, it is possible to produce multiple switches,
as shown in Fig.~3. Such multiple switches are a repeated series of
vortex-antivortex pair creation and annihilation processes. 
In all cases, the total energy is seen to immediately decrease
following each core reversal. However, if the energy provided by the
current pulse after a core reversal is strong enough to overcompensate
the energy dissipation, the energy can still increase before 
further switches occur. A double core switching is obtained with current 
pulses of  about 6$\cdot10^{12}$ A/m$^2$, while triple and quadruple
switches occur with pulses of 7$\cdot10^{12}$ A/m$^2$ and
9$\cdot10^{12}$ A/m$^2$, respectively.  Ultimately, for very large currents
(above 14$\cdot10^{12}$ A/m$^2$), the vortex core is expelled from the
sample.  

The diagram in Fig.~3 (b) shows the number of core switches as a
function of the applied current's strength. Clear ``steps'' are 
observed. While the core reversal mechanism is mediated by the formation of 
a vortex-antivortex pair, it is the annihilation process which leaves the 
vortex with an oppositely-polarized core. As shown in
Ref.~\cite{Hertel06}, such an annihilation is mediated by a magnetic
singularity (Bloch point) which is injected through the sample.  
Since the energy of formation of a Bloch point is uniquely a function of 
the material exchange constant \cite{Tretiakov07}, the annihilation 
(and thus the reversal) process can only occur at specific 
energy values, resulting in the observed steps.

In conclusion, we have presented the possibility of reversing the
polarization of a magnetic vortex core using short current pulses. The
actual magnetization reversal process, which consists of a complicated 
sequence of vortex-antivortex pair creation and annihilation events,
unfolds on a time scale of ca.~40 ps, {\em i.e.} shorter than the
duration of the pulses applied in this study. Further investigations
are required to explore the limits of the operational range for a
controlled, single toggle switching in terms of pulse duration and
pulse  strength, and to determine how short a current pulse can be to
trigger a vortex core reversal. The current-induced vortex core
reversal opens the possibility of addressing individual magnetic
elements in a vortex state within an array of nanoelements. This
feature, in combination with the small size of magnetic vortex cores,
their high thermal stability and the high speed of the reversal
process could make vortex cores interesting candidates for data
storage purposes in future devices.  

Y.L. acknowledges financial support from the Alexander von Humboldt
foundation.

\end{document}